\def \be {\begin{equation}}
\def \eq {\end{equation}}
\def \bee {\begin{eqnarray}}
\def \eqq {\end{eqnarray}}
\def \nn {\nonumber}
\def \bea {\begin{array}{c}}
\def \eqa {\end{array}}

\def \la {\langle}
\def \ra {\rangle}
\def \R {{\bf R}}
\def \C {{\bf C}}
\def \Z {{\bf Z}}
\def \del {\partial}
\def \dels {\partial\kern-.5em / \kern.5em}
\def \As {{A\kern-.5em / \kern.5em}}
\def \Ds {D\kern-.7em / \kern.5em}

\def \a {\alpha}
\def \b {\beta}
\def \dag {\dagger}
\def \g {\gamma}
\def \G {\Gamma}
\def \d {\delta}
\def \eps {\epsilon}
\def \m {\mu}

\def \Lam {\Lambda}
\def \s {\sigma}

\def \one {{\bf 1}}

\def \t {\tau}

\def \II {I\hspace{-.1em}I\hspace{.1em}}

\def \IIB {\mbox{\II B\hspace{.2em}}}

\def \H {{\cal H}}
\def \A {\tilde{A}}
\def \hA {\hat{A}}

\def \tU {\tilde{U}}
\def \U {{\cal U}}
\def \A {{\cal A}}
\def \Gc {{\cal G}}
\def \Gh {\hat{\cal G}}
\def \D {\Delta}
\def \bG {{\bf \Gamma}}

\documentstyle[12pt]{article}

\begin{document}
\begin{titlepage}
\begin{center}
\hfill   UU-HEP/97-06\\
\hfill   JHU-TIPAC-97024 \\
\hfill   IASSNS-HEP-97/140 \\
\hfill   hep-th/9712201 \\

\vskip .5in

{\large Towards a Noncommutative Geometric
Approach\\
to Matrix Compactification}

\vskip .5in

Pei-Ming Ho$^{\dagger}$\footnote{Present Address:
Department of Physics, Jadwin Hall, Princeton University,
Princeton, NJ 08544}, Yi-Yen Wu$^\S$
and Yong-Shi Wu$^{\ddagger}$\footnote{On sabbatical from
Department of Physics, University of Utah, Salt Lake City,
UT 84112-0830}\\
\vskip .3in
{\em $^\dagger$ Department of Physics,
University of Utah \\
Salt Lake City, Utah 84112}
\vskip .3in
{\em $^\S$ Department of Physics and Astronomy,
Johns Hopkins University \\
Baltimore, Maryland 21218}
\vskip .3in
{\em $^\ddagger$ School of Natural Sciences,
Institute for Advanced Study\\
Princeton, New Jersey 08540}

\end{center}

\vskip .5in

\begin{abstract}

In this paper we study generic M(atrix)
theory compactifications that are specified
by a set of quotient conditions. A
procedure is proposed, which both
associates an algebra to each
compactification and leads deductively
to general solutions for the matrix variables.
The notion of noncommutative geometry on
the dual space is central to this construction.
As examples we apply this procedure to various
orbifolds and orientifolds, including ALE spaces
and quotients of tori. While the old solutions
are derived in a uniform way, new solutions
are obtained in several cases.
Our study also leads to a new formulation
of gauge theory on quantum spaces.

\end{abstract}
\end{titlepage}

\newpage
\renewcommand{\thepage}{\arabic{page}}
\setcounter{page}{1}
\setcounter{footnote}{0}

\section{Introduction}

According to the M(atrix) model proposal
\cite{BFSS}, M theory in eleven dimensional
uncompactified spacetime is microscopically
described by the large $N$ limit of the
maximally supersymmetric $U(N)$ Yang-Mills
quantum mechanics. For finite $N$ the model
is conjectured to describe the discrete
light cone quantization of M theory \cite{Suss},
in which one light-cone direction is compactified
on a circle. An attractive feature of M(atrix)
theory is that for the nine transverse directions,
the notion of physical space is a derived one
in the theory. Since the coordinate variables are
valued in the Lie algebra of $U(N)$, the
description of space is novel from the beginning.
\footnote{
Because of supersymmetry, at large distances
the space can be approximately classical \cite{Ban}.}

A well-known generalization of classical (or
commutative) geometry for studying novel
spaces is the noncommutative geometry
pioneered by Connes \cite{Con}. By now it is
known to be relevant to M(atrix) theory at two
different levels. First, a given configuration
of the matrix variables for finite $N$ can be identified
with a regularized membrane \cite{DHN}, whose world
volume is a quantum (or noncommutative) space.
For instance, a regularized spherical membrane
\cite{DHN,KT,Rey} coincides with the quantum sphere
defined in various formulations of noncommutative
geometry \cite{QS}. Interpreted in a different way,
the M(atrix) model action can also be thought of as
describing the dynamics of $N$ D0-branes in the infinite
momentum frame \cite{BFSS}. Previously two of us
\cite{HW} have shown that this action can be understood
as a gauge theory on a discrete noncommutative space
consisting of $N$ points.

Accordingly, at the second level, compactification
in M(atrix) theory is {\it a priori} of noncommutative
nature, since compactification implies
certain specification of allowed background
configurations. The M(atrix) model compactified
on torus and various orbifolds and orientifolds
have already been discussed in the literature.
For toroidal compactifications \cite{BFSS,Tay,GRT},
the original gauge symmetry turns out to give
rise to the usual gauge field theory on a dual
torus, while the winding modes for 1-cycles in the
original compactified space become the momentum
modes in the dual space. Recently in two
interesting papers \cite{CDS,DH}, it was
shown that M(atrix) theory compactification
on a torus can lead to a deformed Yang-Mills
theory on the dual space which is a
quantum torus, and can be interpreted as M theory
configuration with non-vanishing three-form
background on the compactified light-cone and
toroidal directions. This
provides a strong physical motivation for
studying generic M(atrix) compactifications
from the noncommutative point of view.

In this paper we report our recent progress
towards a noncommutative geometric approach
to a wide class of matrix compactifications,
i.e. those on $\cal{M}/\bG$, assuming the
matrix model on a simply connected space
$\cal{M}$ is known, with $\bG$ a
discrete group acting on $\cal{M}$. The
compactification is determined by a set of
quotient conditions, one for each generator
of $\bG$. We will describe a procedure for
solving general solutions to the quotient
conditions. Before doing this, our procedure
naturally associates to each compactification
a noncommutative algebra in which the matrix
variables take values. It starts from here
that the notion of noncommutative geometry
using algebras to describe geometry of
quantum spaces comes into play. Furthermore,
our procedure leads, in a deductive manner,
to solutions which turn out to be
gauge theories on dual quantum space.
We will use several examples to show how
our procedure works in practice. Not only
old solutions, obtained before as classical gauge
theories, are reproduced by our systematic
procedure in a uniform way, new compactification
on quantum spaces is also derived for several
cases, including the Klein bottle, M\"obius
strip and ALE orbifolds. Two different
descriptions \cite{Zwa,KR2} for the dual
space in the case of the Klein bottle were
thought to be in conflict with each other in
the literature; we show that they are both
correct, and a continuous interpolation is
found between them using quantum spaces.

What we obtain corresponds to the ``untwisted''
sector, which may be an anomalous gauge theory
in some cases. We leave for the future the
question about how to derive directly from
M(atrix) model the ``twisted'' sector that
is needed for orbifold and orientifold
compactifications to achieve anomaly
cancellation.

We first reexamine toroidal compactifications in
Sec.\ref{review}, rederiving the results for the
quantum torus with our own procedure. The procedure
for generic M(atrix) compactification is described
in Sec.\ref{procedure}. Then in Sec.\ref{Klein}
and Sec.\ref{CZn}, we will demonstrate how our
procedure works for the Klein bottle \cite{Zwa,KR2}
and the ALE space $\C^2/Z_n$ \cite{DOS,BCD}.
After that some comments on various aspects
of M(atrix) compactifications are made
in subsequent sections.
In the appendix we also consider as examples
$T^2/Z_3$, the finite cylinder \cite{Motl,het,KR1}
and M\"{o}bius strip \cite{Zwa,KR2}.

\section{Toroidal Compactification Revisited}
\label{review}

A $d$ dimensional torus can be defined as
the quotient space $\R^d/\Z^d$, where
$\Z^d$ is generated by $\{c_1,\cdots,c_d\}$
freely acting on $\R^d$ as
\be
c_i: \{x_j\}\rightarrow \{x_j+e_{ij}\},
\eq
where $e_{ij}$ define a $d$ dimensional
lattice in $\R^d$. The toroidal
compactification is defined by the
quotient conditions
\cite{BFSS,Tay,GRT}:
\be \label{torus}
U_i^{\dagger}X_j U_i=X_j+e_{ij}, \quad i,j=1,\cdots,d.
\eq

Standing as a fundamental theory, M(atrix)
theory itself should contain the answer
for all compactifications described by
relations of this type. Although a complete
answer including ``twisted'' sectors is not
yet generally known to us, as a first step
in this paper we will try to solve these
equations for the ``untwisted'' sector,
completely inside the framework of the theory.

One may choose an infinite dimensional
matrix representation for the $U_i$'s in
eq.(\ref{torus}), motivated by physical
considerations. In our treatment, we prefer
to think of them as algebraic elements
tensored with an $N\times N$ unit matrix.
\footnote{The $e_{ij}$'s on the right hand
side is understood as proportional to the
unity in the algebra tensored with
the $N\times N$ unit matrix.}

To find the underlying algebra
for the $U_i$'s, we note that
eq.(\ref{torus}) implies that
\be
U_j U_i U_j^{\dag}U_i^{\dag}X_k
U_i U_j U_i^{\dag}U_j^{\dag}=X_k.
\eq
For toroidal compactifications, we
should not have any additional constraints
other than those in eq.(\ref{torus}).
Therefore, if we assume that the only
central elements in the algebra of the $U_i$'s
are constant times the unity $\one$,
we are allowed to impose the following
constraints:
\be \label{UUUU}
U_i U_j U_i^{\dag}U_j^{\dag}=q_{ij}\one,
\eq
or equivalently,
\be \label{U1U2}
U_i U_j=q_{ij} U_j U_i
\eq
with $q_{ij}$ certain phase factors.
Different choices of these phases may
lead to different solutions, implying
that compactification is not completely
fixed by the quotient conditions.

The algebra (\ref{U1U2}) is the same as
the algebra of a quantum torus \cite{QT}.
For $d=2$ the algebra (\ref{U1U2}) has
an $SL(2,\Z)$ symmetry
\be \label{SL2Z}
U_1\rightarrow (U_1)^a\, (U_2)^b,\quad
U_2\rightarrow (U_1)^c\, (U_2)^d,
\eq
where $a,b,c,d$ are the entries of an
$SL(2,\Z)$ matrix. It was first pointed
out in \cite{CDS,DH} that the phase factors
$q_{ij}$ can be related to M theory
compactification with non-zero background
three-form field in the compactified null
and toroidal directions.

{}From the point of view of the covering
space, $U_i$'s are translation operators,
so it is natural to write for $N=1$:
$X_j=e_{ij}\tilde{\s}_i$ and $U_i=\exp(-\tilde{D}_i)$,
where $\tilde{D}_i=\frac{\del}{\del\tilde{\s}_i}+i\tilde{A}_i$
is the covariant derivative for a $U(1)$ gauge field.
By a Fourier transform, the solution in the dual
space is \cite{BFSS,Tay,GRT}
\bee
&U_i=e^{i\s_i}, \quad X_j=-ie_{ij}D_i, \label{XU2}\\
&D_i=\frac{\del}{\del\s_i}+iA_i(\s).
\eqq
In this dual representation, the solution can
easily be generalized to $N\times N$ matrices.
This is the type of solutions we are looking
for in the context of M(atrix) compactifications.
New physical degrees of freedom reside in $X$,
while the $U$'s are fixed algebraic elements.

\subsection{Classical Torus}
\label{class-torus}

First we review the commutative case
$q_{ij}=1$ \cite{BFSS,Tay,GRT}. In this case, the
algebra of the $U_i$'s is commutative, and
now they are viewed as coordinate functions
on the dual (ordinary) torus parametrized
by $\s_i$, and the $X$'s as covariant
derivatives. Mathematically eq.(\ref{XU2})
is the general solution of (\ref{torus}),
with the $X$'s and $U$'s being elements in
the product of the algebra of differential
calculus on a torus and the algebra of
$N\times N$ matrices. Physically, the
M(atrix) theory compactified on $T^d$
for $d\leq 3$ is the $d+1$ dimensional
SYM on the dual torus \cite{BFSS}.

Comparing with the uncompactified
M(atrix) theory, we are adjoining the
new elements ${\del}/{\del\s_i}$ and
$\exp(i\s_i)$ to the algebra of
$N\times N$ matrices for the
compactification on a torus.
The reason why we are allowed to
adjoin these new algebraic elements
is that the compactification on a torus
introduces new dynamical degrees of
freedom corresponding to the winding
string modes that are not present
in the uncompactified theory.
In general, for compactification on
different spaces we need to adjoin
different new elements to
the algebra of $N\times N$ matrices.

\subsection{Quantum Torus}
\label{quantum-torus}

For $q_{ij}\neq 1$, we need to find
out the new elements to be adjoined to
the algebra of compactification.
To define the algebra and to solve for $X$
in this noncommutative case, we first define
an auxiliary Hilbert space $\H$, on which
the $U_i$'s are represented as operators:
It by definition consists of the ``vacuum'',
denoted by $\ra$, as well as states obtained
by acting polynomials of $U_i$'s on the vacuum.
For $d=2$ the symmetry (\ref{SL2Z}) induces
an $SL(2,\Z)$ symmetry on the Hilbert space.
This is the S duality of type $\IIB$ theory.

The Hilbert space $\H$ is spanned by the states
$\{\,U_1^{m_1}\cdots U_d^{m_d}\ra\,\}$ with $m_i\in\Z$.
This Hilbert space is different from those
introduced in \cite{CDS}.\footnote{In the notation
of \cite{CDS} our $\H$ superficially corresponds
to the case with $p=1, q=0$, but $p/q$ appears in
some of the relations given by them.} For later convenience,
we define a set of operators $\del_i$ by
\be \label{Mi}
\del_i U_1^{m_1}\cdots U_d^{m_d}\ra
=m_i U_1^{m_1}\cdots U_d^{m_d}\ra.
\eq
It follows that
\be
\del_i U_j=U_j(\del_i+\d_{ij}).
\eq
Thus $\del_i$ is the (quantum) derivative
with respect to the exponent of $U_i$.

The inner product on $\H$ should be invariant under
the group $\Gh(\A)$ of gauge transformations of the $U_i$'s
which preserve the quotient conditions (\ref{torus}).
Since $X_i$ is generic,
the only possible such transformation is
\be
U_i\rightarrow g_i^{\dag}U_i g_i=e^{i\phi_i}U_i,
\eq
where $g_i=\exp(-i\phi_i\del_i)$.
This implies that the inner product is defined by
$\la f|g\ra=\la f^{\dag}g\ra$,
where $f,g$ are functions of $U$ and
\be
\la U_i^{m_i}\cdots U_d^{m_d}\ra = \d^{m_1}_0\cdots\d^{m_d}_0
\eq
up to normalization.
Note that the vacuum expectation value $\la\cdot\ra$
happens to equal to the trace over the Hilbert space,
which can be determined directly
by requiring that it has the property of cyclicity:
$\la fg\ra=\la gf\ra$ for any two functions of $U$.
By a Fourier transform on the basis:
$|\s\ra=\sum_n\exp(in_i\s_i)U_1^{n_1}\cdots U_d^{n_d}\ra$,
where $\s=(\s_1,\cdots,\s_d)$ and $n=(n_1,\cdots,n_d)$,
the trace on $\H$ turns into the integration on
a $d$-torus parametrized by $\s$.
The integration on a quantum torus can be independently
defined with respect to the $\Gh(\A)$-invariant measure
$\Pi_i U_i^{\dag}dU_i$ by using Stoke's theorem.

Let the action of $X_j$ on the vacuum be given by
\be \label{XA}
X_j\ra=\hA_j(U)\ra,
\eq
where $\hA_j$ is a function of the $U_i$'s.
Using (\ref{torus}) and (\ref{XA}),
we can calculate the action of $X_j$ on any state:
\bee
X_j U_1^{m_1}\cdots U_d^{m_d}\ra&=&
U_1^{m_1}\cdots U_d^{m_d}(e_{ij}m_i+\hA_j)\ra \nn\\
&=&(e_{ij}\del_i+A_j)U_1^{m_1}\cdots U_d^{m_d}\ra;
\eqq
i.e. in general $X_j= e_{ij}\del_j + A_j$, where
$A_j$ are functions of $\tU_i=(\Pi_{j\neq i}
q_{ji}^{\del_j})U_i$, obtained by replacing
$U_i$'s in $\hA_j(U)$ with $\tU_i$'s
and reversing the ordering of a product.

The solution of the $X_i$'s are functions of
operators commuting with all $U$'s, i.e.
$\tU_i U_j=U_j\tU_i$ for all $i,j$.
The commutation relations among the $\tU$'s are given by
\be
\tU_i\tU_j=q_{ij}^{-1}\tU_j\tU_i.
\eq
This is just the algebra for a quantum torus related
to that of $U$ by a transformation
$q_{ij}\rightarrow q_{ij}^{-1}$ \cite{CDS}.
The Hilbert space is also spanned by
$\{\tU_1^{m_1}\cdots\tU_d^{m_d}\ra\}$,
and the operators $\del_j$ act on $\tU_i$
in the same way as they act on $U_i$.
It is thus natural to think that
$X_j$ are the covariant derivatives
on the dual quantum torus given by $\tU_i$.

The same result was obtained in \cite{CDS}
in a different way. They noticed that a
generic solution of (\ref{torus}) is composed
of a special solution and a homogeneous solution,
and that homogeneous solutions are the elements
in the algebra commuting with all the $U_i$'s.
Also, they used a Hilbert space different from ours.
While the set of $U$-commuting elements may be
found by brute force when the algebra is given,
we see that they automatically arise in our procedure.
For a different compactification associated
with another set of quotient conditions, the trick
of using $U$-commuting operators may no longer work,
but we will demonstrate below that the same
procedure we used above always works.

Let us now make a remark about the gauge field $A_i$.
As in usual gauge theories,
the gauge field $A_i$ does not have to be a well-defined
function on the dual quantum torus.
Without going into details about the notion of
principal bundle and connection on quantum spaces \cite{Con},
we simply say that the requirement on $A_i$
is that all quantities be invariant under
\be \label{Xe}
X_j\rightarrow X_j+e_{ij}
\eq
are well defined.
For instance $(-i\log U_i)$ is only defined
up to $2n\pi$. Yet $A_j=-im_i\log(U_i)e_{ij}$
with integers $m_i$ is acceptable, because the
ambiguity in its value matches precisely the
gauge transformation (\ref{Xe}). In fact
these are the configurations of D-branes
wrapping on the torus.

\section{Generic Compactification}
\label{procedure}

Consider the compactification of M(atrix) theory
on the quotient space $\cal{M}/\bG$,
\footnote{In fact we should consider the
quotient of a superspace in order to include
the fermionic part from the beginning.}
where $\cal{M}$ is a simply connected space
($\pi_1(\cal{M})=\one$) on which the M(atrix)
theory is known, and $\bG$ is a discrete group
acting on $\cal{M}$. If $\bG$ acts freely, it
is the fundamental group of the compactified
space.

Denote the action of $c\in \bG$ on $\cal{M}$
by $\Phi(c)$. Then the compactified M(atrix)
theory is obtained by imposing the following
constraints: For each element $c\in \bG$,
\be \label{patch0}
U(c)^{\dag}X_a U(c)=\Phi_a(c)(X),
\eq
where $X_a$ represent all M(atrix) theory
variables $A_0, X_i$ and $\Psi$. If $\bG$
is generated by a set of elements $\{c_i\}$,
one may only need to write down such
relations for each generator $c_i$.
We will call these relations
``quotient conditions''.

For orientifolds, the group $\bG$ is endowed with
a $\Z_2$-grading: We associate a number $n(c)=0,1$
to each element $c\in\bG$, and if $c_1 c_2=c_3$ then
$n(c_1)+n(c_2)=n(c_3)$ (mod $2$).
The quotient condition (\ref{patch0}) is
generalized to
\be \label{patch}
U(c)^{\dag}X_a U(c)=\left\{\begin{array}{lll}
   \Phi_a(c)(X) & \mbox{if} & n(c)=0\\
   (\Phi_a(c)(X))^* & \mbox{if} & n(c)=1.
   \end{array}\right.
\eq
Here the complex conjugation $*$ corresponds to
the transpose for Hermitian matrices
$X$, which implies orientation reversal of open
strings stretched between D0-branes.

The quotient conditions have to be
consistent with the action. Since the action
of M(atrix) theory is invariant under
gauge transformations: $X\rightarrow U^{\dag}XU$,
the quotient conditions are consistent
with the action only if the action is
also invariant under the transformations:
\be \label{gauge}
X_a\rightarrow \Phi_a(c)(X)
\eq
for all $c\in\bG$.
A function of the $X$'s and their time
derivatives is a gauge invariant physical
observable if it is invariant under (\ref{gauge}).

We will give below a procedure for solving
relations of the type (\ref{patch0}) or
(\ref{patch}). By this we mean that we shall
define the algebra $\A$ in which the relations
are understood, and then find the most general
solution of $X_a$ as algebraic elements
in the algebra $\A$. The physical degrees
of freedom of the $X_a$'s reside in the moduli
of the solutions to the quotient conditions.

To define the algebra $\A$, first we note
that all the $U$'s are considered
as fixed elements in $\A$. They form a
subalgebra of $\A$ which is constrained
by the quotient conditions by requiring that
the quotient conditions exhaust all desired
constraints on $X$. If there is a relation
$c_1 c_2\cdots c_n=\one$ in the group $\bG$,
from the quotient conditions for these $c$'s,
we will get equations of the form:
\be \label{PU}
P(U)^{\dag}X P(U)=X \quad \mbox{for all $X$'s,}
\eq
where $P(U)=U(c_1)\,U(c_2)\cdots U(c_n)$ is the
corresponding product of the $U$'s. This relation
would impose a new constraint on $X$
unless
\be\label{22}
P(U)=q\one,
\eq
where $q$ is a phase factor.
For orientifolds, let $C$ denote the complex
conjugation operator:
\be
CaC=a^*
\eq
for all $a\in\A$. We have $C^{\dag}=C$ and
$C^2=\one$. Eq.(\ref{patch}) is then
equivalent to
\be
R(c)^{\dag}X_a R(c)=\Phi_a(c)(X),
\eq
where $R(c)=U(c)C^{n(c)}$. So eq.(\ref{22})
is replaced by $P(R)=q\one$. We define the
algebra of $U$, called the $U$-algebra,
by imposing all such relations. We can view
non-orientifolds as the special case
with $n(c)=0$ for all $c\in\bG$.

It can be shown \cite{Work} that these relations
can be characterized by a faithful projective
representation of $\bG$. Following \cite{CDS,DH},
it is natural to suggest that the
cohomologically invariant phases in a
nontrivial 2-cocycle on $\bG$ associated with
the projective representation correspond
to a nontrivial background field on the
compactified space. Accordingly, compactification
defined by the quotient conditions is completely
characterized by projective representations of
the group $\bG$, and the moduli space of the
$U$-algebra (more precisely, the space of
cohomologically invariant $q$-parameters in
a 2-cocycle) may correspond to
part of the moduli of M theory compactifications.
We take this as a strong motivation for studying
M(atrix) theory compactification with
nontrivial 2-cocycles.

Knowing the $U$-algebra, we can construct a
Hilbert space $\H$ to represent it, which
consists of a ``vacuum'' denoted by $\ra$ and
all polynomials of the $R(c)$'s acting on the
vacuum.
The algebra $\A$ is then defined as the
tensor product of the algebra of operators on
$\H$ with the algebra of $N\times N$ matrices.
In the action of M(atrix) theory, the total
trace is now composed of the trace over $\H$
and the trace over $N\times N$ matrices.

Physically the states in $\H$ correspond to string
modes winding around noncontractible 1-cycles
in the compactified space associated with
elements in the group $\bG$. By adjoining
this Hilbert space to the space of $N$-vectors
on which the algebra of $N\times N$ matrices is
represented, we take care of the new string
winding modes arising from the compactification.

For a given algebra $\A$ we define
the unitary group $\U(\A)$ to be the
group of all unitary elements in $\A$.
Let $\Gc(\A)$ be the subgroup of $\U(\A)$
which preserves the quotient conditions, i.e.
\be
R(c)^{\dag}g^{\dag}X_a g R(c)=\Phi_a(c)(g^{\dag}X g)
\eq
for all $g\in\Gc(\A)$.
$\Gc(\A)$ can be viewed as the group
of gauge transformations on $X$:
\footnote{
In fact $\Gc(\A)$ contains more than what
we usually call a gauge group on the dual space
for it also contains the translation group $\Gh(\A)$
to be introduced below.}
\be
X_a\rightarrow g^{\dag}X_a g
\eq
which survives the compactification.
As it was shown in the previous section,
the definition of the dual space
may be inferred from the gauge field,
or equivalently from the gauge group $\Gc(\A)$.
In general the compactified M(atrix) theory may not
be identified with a traditional gauge theory
on a classical manifold.
We will consider this as a natural generalization of
the notion of gauge theories.

On the other hand, $\Gc(\A)$ induces
a group of transformations on $R(c)$, denoted by $\Gh(\A)$:
\be
R(c)\rightarrow g R(c) g^{\dag}, \quad g\in\Gc(\A),
\eq
which preserve the quotient conditions (\ref{patch}).
Because we shall allow the most general solution of $X$,
the only possible transformation on $R(c)$ is
to multiply them by certain phase factors,
and thus the group $\Gh(\A)$ is an Abelian group.
Since different choices of the $R(c)$'s related
by $\Gh(\A)$ are equivalent by a gauge transformation,
the compactification should be invariant under $\Gh(\A)$.
Roughly speaking, $\Gh(\A)$ is the translation group of
the dual space.

The prescription for deriving the general
solution for $X$ in the algebra $\A$
was first invented by Zumino \cite{Zum} to study
problems in quantum differential calculi.
(Mathematically these two problems are similar in nature.)
The prescription is:

\begin{enumerate}

\item \label {step1}
As mentioned above, we define a Hilbert space $\H$
consisting of all polynomials of the $R(c)$'s acting on the vacuum.
The inner product on $\H$ has to be fixed
to respect the symmetry group $\Gh(\A)$.
The algebra $\A$ is defined to be the tensor product of
the algebra of operators on $\H$ with
the algebra of $N\times N$ matrices.

\item \label{step2} Require the $X_a$'s be
operators acting on $\H$ and write the action
of $X_a$ on the vacuum as
\be \label{Xvac}
X_a\ra=\hA_a(R)\ra,
\eq
where $\hA_a(R)$ is a function of the $R(c)$'s.
All physical degrees of freedom of $X_a$
reside in $\hA_a$, which gives the generalized
gauge field. The action of $X_a$ on an arbitrary
basis state can be obtained by using the quotient
conditions to commute $X_a$ through the $R(c)$'s
until it reaches the vacuum and then using
eq.(\ref{Xvac}).

\item \label{step3}
To find an explicit expression for $X_a$,
\footnote{It is not necessary to have an
explicit expression of $X_a$ in terms of other
operators as long as $X_a$ is already
well defined as an operator on $\H$
as in step \ref{step2}.
But it can be helpful in studying the model.}
one needs to find a set of convenient
operators on $\H$. The type of operators
(\ref{Mi}) used for toroidal compactification
are often very useful. As we did in Sec.
\ref{quantum-torus}, to write $X_a$ as a
function of $\del_i$ and $U_i$, one needs to
find the action of $X_a$ on a state
$U_1^{m_1}\cdots U_d^{m_d}\ra$ as a
function $F(m_1,\cdots,m_d;U)$ acting on the
state. Then we can replace $F$ by another
function $\tilde{F}$ of $\del$ and $U$.

\end{enumerate}

To gain some insight of the compactified
theory, we note that in general we may view
the resulting theory as a (deformed) gauge
field theory on a dual quantum space. In the
spirit of noncommutative geometry, the
$U$-algebra can be viewed as the algebra of
functions on the dual quantum space.
In addition one may follow the standard
procedure used in the study of quantum
differential calculus on quantum spaces with
quantum group symmetry \cite{DC}
\footnote{In our problem the symmetry group is
$\Gh(\A)$, which is just a classical group.
But they play similar roles in this formulation.}
to define a deformed differential calculus on
the $U$-algebra. Once the derivatives
(such as the $\del_i$ in the previous section)
on the dual quantum space are defined, we can use
them to express $X_a$ and see that the bosonic
$X_i$'s can be thought of as covariant derivatives.
In other words, the present approach can be
directly used to define gauge theory on a quantum
space and is different from most other
existing approaches to defining them
in the following sense: Given the algebra of
functions on a quantum space, usually one will define
the gauge field to be a function on the quantum space,
but in general our procedure gives a gauge field
as an operator, for instance a pseudo-differential
operator, on the quantum space.

We will demonstrate below how our above procedure
works, for example, for the compactification
on the orientifold of Klein bottle and the ALE
space of $\C^2/\Z_n$. In the appendix, we will also
apply the prescription to the following
orbifolds and orientifolds: $T^2/\Z_3$, cylinder
($S^1\times S^1/\Z_2$) and M\"{o}bius strip.

\section{Klein Bottle} \label{Klein}

The Klein bottle can be defined as $\R^2/\bG$,
where $\bG$ acts on $\R^2$ by
\bee
&c_1: (x_1, x_2)\rightarrow (x_1+2\pi R_1, x_2), \\
&c_2: (x_1, x_2)\rightarrow (-x_1, x_2+\pi R_2).
\eqq
The group $\bG$ is generated by $c_1,c_2$
with the commutation relation:
\be \label{c1c2}
c_1 c_2 c_1 c_2^{-1}=\one.
\eq
As an orientifold, its $\Z_2$-grading is defined by
$n(c_1)=0$ and $n(c_2)=1$.

Thus the quotient conditions are \cite{Zwa,KR2}
\bee
U_i^{\dag}X_j U_i&=&X_j+2\pi\d_{ij}R_j,
\quad i,j=1,2, \label{T2}\\
U_3^{\dag}X_1 U_3&=&-X_1^*, \label{X1}\\
U_3^{\dag}X_2 U_3&=&X_2^*+\pi R_2, \label{X2}
\eqq
where $U_1=U(c_1)$, $U_2=U(c_2^2)$ and $U_3=U(c_2)$.
Note that since $X$'s are Hermitian, we have $X^T=X^*$.
The conditions for $U_2$ are
direct results of (\ref{X1}),(\ref{X2}).

Since $R(c)=U(c)C^{n(c)}$, it is
easy to verify that the following relations
are compatible with the quotient conditions
(\ref{T2})-(\ref{X2}):
\bee
U_1 U_2&=&q_{12}U_2 U_1, \label{U12}\\
U_1 U_3&=&q_{13}U_3 U_1^T, \label{U13T}\\
U_2 U_3&=&q_{23}U_3 U_2^*, \label{U23}\\
U_3 U_3^*&=&q_3 U_2. \label{U33*}
\eqq
We shall rescale $U_2$ to set $q_3=1$.
Using (\ref{U33*}) we find $q_{23}=1$ from (\ref{U23}).
Consistency also requires that $q_{12}=q_{13}^2$.
We will denote $q_{13}$ by $q$. (So the projective
representations of the group $\bG$ are labelled
only by a phase factor $q$.)

We will see below that the case studied
in \cite{Zwa} corresponds to the case $q=1$
where the dual space is a cylinder, and
the case studied in \cite{KR2} corresponds to
the case $q=-1$ where the dual space is a
Klein bottle. We have obtained a one-parameter
moduli for this compactification.

The Hilbert space $\H$ is defined to be
$\H=\{U_1^m(U_3 C)^n\ra| m,n\in\Z\}$,
or equivalently
$\{U_1^m U_2^n\ra, U_1^m U_2^n U_3 C\ra| m,n\in\Z\}$.
We define some operators for later convenience:
\bee
&\del_1 U_1^m(U_3 C)^n\ra=m U_1^m(U_3 C)^n\ra, \\
&\del_2 U_1^m U_2^n(U_3 C)^s\ra=n U_1^m U_2^n(U_3 C)^s\ra, \\
&K U_1^m(U_3 C)^n\ra=U_1^m(U_3 C)^{n+1}\ra, \\
&\eps U_1^m(U_3 C)^n\ra=(-1)^n U_1^m(U_3 C)^n\ra,
\eqq
where $m,n\in\Z$ and $s=0,1$.
It follows that $\del_1, \del_2$ acts on
$U_1, U_2$ as derivatives.
The commutation relations between the derivatives
and functions can easily be derived.

Following the prescription described in the last section,
we see that the solution is of the form of a gauge field
\bee
&X_1=2\pi R_1\del_1+\frac{1}{2}\hA_1(q^{-N}U_1,K)(1+\eps)
-\frac{1}{2}\hA_1^*(q^N U_1^{-1},K)(1-\eps), \\
&X_2=\pi R_2 N+\frac{1}{2}\hA_2(q^{-N}U_1,K)(1+\eps)
+\frac{1}{2}\hA_2^*(q^N U_1^{-1},K)(1-\eps),
\eqq
where $N=2\del_2+(1-\eps)/2$ acts on $\H$ by
\be
N U_1^m(U_3 C)^n\ra=n U_1^m(U_3 C)^n\ra.
\eq

While the Klein bottle is a quotient of the torus,
we will see below that
the compactification on the former is a gauge theory
on a quotient of the dual torus for the latter.
We have
\be
X_1=2\pi R_1\del_1+A_1, \quad
X_2=2\pi R_2(\del_2+\frac{1-\eps}{4})+A_2.
\eq
The gauge fields are given by
\be
A_i=\frac{1}{2}(A_{i0}+A_{i1}K)(1+\eps)
+(-1)^i\frac{1}{2}(B_{i0}+B_{i1}K)(1-\eps),
\eq
where $A_{ij}$ and $B_{ij}$ ($i=1,2$ and $j=0,1$)
are functions of $\tU_1, \tU_2$
with $\tU_1=q^{-2\del_2}U_1$ and $\tU_2=q^{2\del_1}U_2$
satisfying the algebra of the dual torus
\be \label{U-dual}
\tU_1\tU_2=q^{-2}\tU_2\tU_1.
\eq
It is
\be \label{Ai}
A_{ij}(\s_1-h,-\s_2)=B_{ij}^*(\s_1,\s_2), \quad i=1,2,\; j=0,1,
\eq
where $q=\exp(ih)$, $\tU_1=\exp(i\s_1)$ and $\tU_2=\exp(i\s_2)$.
It can be checked that
\be
U_3^{\dag}A_i U_3=(-1)^i A_i^*, \quad i=1,2,
\eq
and all quotient conditions are automatically satisfied.

The condition (\ref{Ai}) relates $A_i(\s_1^*+h,-\s_2^*)$
to $A_i(\s_1,\s_2)^*$, which is a function of $(\s_1^*,\s_2^*)$.
So if the value of $A_i$ at $(\s_1+h,\s_2)$ is known,
then its value at $(\s_1,-\s_2)$ is fixed.
If $q=\exp(i2\pi/(2k))$ for an integer $k$,
the fundamental region on which the values of $A_i$
can be freely assigned is a Klein bottle
of area $(2\pi)^2/(2k)$.
If $q=\exp(i2\pi/(2k+1))$,
the fundamental region is a cylinder
of area $(2\pi)^2/2(2k+1)$.
In particular, for $q=1$ it is a cylinder,
and for $q=-1$ it is a Klein bottle.
It was argued in \cite{KR2,KR1} that only the latter case
gives the area-preserving diffeomorphism group
as the gauge group of the model in the large $N$ limit.
The gauge group in the bulk of the fundamental region
is $U(2N)$ and the gauge group on fixed points
of the map $(\s_1,\s_2)\rightarrow(\s_1-h,-\s_2)$
is $O(2N)$.

$K$ and $\eps$ can be represented by $2\times 2$ matrices. Let
\bee
&K=e^{i\s_2/2}\left(\begin{array}{cc}
 0 & 1 \\ 1 & 0 \end{array}\right), \quad
\eps=\t_3=\left(\begin{array}{cc}
 1 & 0 \\ 0 & -1 \end{array}\right), \\
&A_i=\left(\begin{array}{cc}
 \a_i & \b_i \\ \g_i & \d_i \end{array}\right)
=\left(\begin{array}{cc}
 A_{i0} & (-1)^i B_{i1}e^{i\s_2/2} \\
 A_{i1}e^{i\s_2/2} & (-1)^i B_{i0} \end{array}\right).
\eqq
The results above can then be rewritten as
\be
\left.\left(\begin{array}{cc}
 \d_i & \g_i \\ \b_i & \a_i \end{array}\right)
\right|_{(\s_1+h,\s_2)}=(-1)^i
\left.\left(\begin{array}{cc}
 \a_i^* & \b_i^* \\ \g_i^* & \d_i^* \end{array}\right)
\right|_{(\s_1,-\s_2)}.
\eq

The $2\times 2$ unit matrix and $K$ (for ``fixed'' $\s_2$)
generate the algebra of functions on $\Z_2$,
and $\eps$ is a derivative on $\Z_2$
in the sense of noncommutative geometry \cite{Con}.
Thus $X_i$ can be viewed as covariant derivatives
on the dual space which is the product of
a classical space parametrized by $\s_1, \s_2$
and a quantum space of two points ($\Z_2$).
The Hilbert space can also be written as a column
of two functions of $\tU_1$ and $\tU_2$.
Thus it is natural to say that the dual space
has coordinates $\tU_1$, $\tU_2$ and $K$,
where $\tU_i$ satisfy the same algebra as $U_i$ ($i=1,2$)
with $q\rightarrow q^{-1}$.

The trace over $\H$ is equivalent to the composition
of the integration over $(\s_1,\s_2)$
and the trace over the $2\times 2$ representation of $K$ and $\eps$.
The integration has the cyclicity property
so that the M(atrix) theory action is gauge invariant.

As it was noted in \cite{CDS},
the algebra of the dual quantum torus (\ref{U-dual})
can be realized on functions on a classical torus as
the star product:
\be
(f\ast g)(\s)=\left. q^{\del_2\del'_1-\del_1\del'_2}
f(\s)g(\s')\right|_{\s'=\s}.
\eq
Therefore the action of M(atrix) theory appears to be
the action for a field theory defined on $T^2$ with
higher derivative terms. It is yet to be studied
how to make sense of such theories.

As a side remark we note that
the calculation above can be done with a little more ease
if we impose the reality conditions
$U_1^*=U_1^{-1}$, $U_2^*=U_2$ and $U_3^*=U_3$,
which are consistent with the $U$-algebra.
The result is independent of such conditions.

So far we have ignored the transverse bosonic and fermionic
fields in the M(atrix) theory.
The quotient conditions on them are \cite{Zwa,KR2}
\bee
U_i^{\dag} A_0 U_i=A_0, & U_3^{\dag} A_0 U_3=-A_0^*,\\
U_i^{\dag} X_a U_i=X_a, & U_3^{\dag} X_a U_3=X_a^*,\\
U_i^{\dag}\Psi U_i=\Psi, & U_3^{\dag}\Psi U_3=\G_{01}\Psi^*,
\eqq
where $i=1,2$, $a=3,\cdots,9$, and
$\Psi$ is in the Majorana representation.
It is straightforward to solve these relations in the same way.
These quotient conditions can be determined by
required surviving SUSY or
by their consistency with the M(atrix) theory Lagrangian \cite{BFSS}:
\be \label{Lag}
L=\mbox{Tr}\left(\frac{1}{2}(D_0 X_i)^2+\frac{1}{4}[X_i,X_j]^2
-\frac{1}{2}\Psi^{\dag}D_0\Psi
-\frac{1}{2}\bar{\Psi}\G^i[X_i,\Psi]\right),
\eq
where $D_0=\frac{\del}{\del t}+iA_0$.

The dynamical SUSY transformation of M(atrix) theory is \cite{BFSS}
\bee \label{dyn-SUSY}
\d X_{\mu}&=&i\bar{\eps}\G_{\mu}\Psi, \quad \mu=0,\cdots,9, \\
\d\Psi&=&(D_0 X_i)\G^{0i}\eps+\frac{i}{2}[X_i,X_j]\G^{ij}\eps,
\quad i,j=1,2,\cdots,9;
\eqq
and the kinetic SUSY transformation is
\be \label{kin-SUSY}
\tilde{\d}X_{\mu}=0, \quad \tilde{\d}\Psi=\tilde{\eps}.
\eq
One half of the dynamical SUSY is preserved
by the compactification on a Klein bottle.

\section{$\C^2/\Z_n$} \label{CZn}

The quotient condition for $\C^2/\Z_n$ is
\be
U^{\dag}Z_a U=qZ_a, \quad a=1,2,
\eq
where $Z_1=X_1+iX_2$, $Z_2=X_3+iX_4$ and $q=\exp(2\pi i/n)$.
It follows that $U^{-n}Z_a U^n=Z_a$.
Following our procedure,
the $U$-algebra is given by $U^n=p\one$,
where $p$ is a phase.
Rescaling $U$ by $p^{1/n}$, we find
\be \label{Un}
U^n=1.
\eq

The Hilbert space is $\H=\{U^m\ra| m=0,1,\cdots,n-1\}$.
Let $Z_a\ra=A_a(U)\ra$,
where $A_a(U)=\sum_{m=0}^{n-1}\a_{am} U^m$.
The action of $Z$ on $\H$ is
\bee
Z_a U^m\ra&=&q^m U^m A_a\ra \\
&=&A_a(U)q^M U^m\ra,
\eqq
where $M$ is defined by $M U^m\ra=m U^m\ra$.
The solution of $Z_a$ is thus $Z_a=A_a(U)q^M$.
Instead of $M$, one can also use $V$ defined by
$UV=qVU$ and $V\ra=\;\ra$.
Thus $Z$ can also be expressed as
\be
Z_a=A_a(U)V^{-1}.
\eq

$U$ and $V$ can be realized as $n\times n$ matrices:
\be \label{UV}
U_{ij}=\d_{i,(j-1)}, \quad V_{ij}=q^i\d_{ij},
\eq
where $U_{ij}$ is non-vanishing only if $i=j-1$ (mod $n$). We find
\be \label{Z}
(Z_a)_{ij}=\sum_m \a_{am}q^{-j}\d_{i,(j-m)}, \quad i,j=0,1,\cdots,n-1.
\eq
This is exactly what one would expect through
the same line of reasoning Taylor used \cite{Tay}
for toroidal compactifications.
The coefficient $a_m$ represents the string
stretched between D0-branes separated by $m$
copies of the fundamental region.

In the representation (\ref{UV}),
$U$ is viewed as an operator that
shifts one point in $\Z_n$ to the next point.
In a dual representation where $U_{ij}=q^{-i}\d_{ij}$,
$U$ can be viewed as the generator
of the algebra of functions
on the dual quantum space $\Z_n$,
and $V$ becomes the shift operator.
Thus we see that the dual of $\Z_n$ is also $\Z_n$.

The group $\Gc(\A)$ is generated by $U$ and $V$.
A unitary function $g(U)$ induces a gauge transformation
$A(U)\rightarrow g^{\dag}(U)A(U)g(qU)$.
In the dual representation where $U$ is diagonal,
it is easy to see that the gauge group of this theory is $U(N)^n$.
The fields $A_a$ are now diagonal blocks of $N\times N$ matrices
with each block transforming
in the fundamental and antifundamental representations
under two adjacent $U(N)$ factors \cite{BCD}.

The gauge transformation by $V^k$ is
$A(U)\rightarrow A(q^k U)$,
which is in fact a translation (cyclic permutation)
on the dual space $\Z_n$.
This also corresponds to the only nontrivial elements
in $\Gh(\A)$: $U\rightarrow q^k U$.
Requiring its invariance under $\Gh(\A)$,
the inner product on $\H$ is fixed to be
$\la U^k\ra=\d^k_0$ for $k=0,1,\cdots,n-1$.

Note that in M(atrix) theory it is only
the field strength defined by $[X_i,X_j]$ (for flat space)
and other gauge invariant quantities
that need to be well defined on the dual space.
For instance, $U^{1/n}$ is only defined up to
an integral power of $q$.
But it is acceptable to have $A(U)=U^{m/n}F(U)$ with
$m=0,1,\cdots,n-1$, where $F(U)$ is a polynomial of $U$.
The reason is that this ambiguity is precisely of
the form of a gauge transformation on $X$
so all gauge invariant quantities are still well defined.

Denote $X_0=A_0$.
The rest of the quotient conditions are
\bee
U^{\dag}X_{\mu}U&=&X_{\mu}, \quad \mu=0,5,\cdots,9, \\
U^{\dag}\Psi U&=&\Lam\Psi,
\eqq
where $\Lam=\exp(-\pi(\G^{12}+\G^{34})/n)$.
Because $\Lam^n=-\one$, eq.(\ref{Un})
should be replaced by $U^n=(-1)^F$,
where $F$ is the fermion number operator.
It is easy to see that $A_0$, $X_{\m}$ and $\Psi$
are in the adjoint representation of $U(N)^n$.

It is easy to see that the quotient
conditions for $\C^2/\Z_n$
preserve one half of the dynamical SUSY
and one half of the kinetic SUSY.

\section{Noncommutative Geometry and T Duality}
\label{T-dual}

Let us recall how the notion of
noncommutative geometry naturally arises
as a generalization of classical geometry.
We know that if a classical space is given,
one can immediately define the algebra
of functions on that space.
According to the Gelfand-Naimark theorem,
the converse is also true:
any commutative $C^*$ algebra is isomorphic
to the algebra of functions (vanishing at infinity)
on a locally compact Hausdorff space,
which can be constructed as the space of
maximal ideals of the algebra.
The notion of the algebra of functions and
that of the underlying space are dual to
each other via the Gelfand map.
This motivates the generalization of
classical spaces to quantum spaces.
A quantum space is simply defined
as the underlying space of a noncommutative algebra.

The dual space for a M(atrix) compactification
can thus be roughly viewed as the underlying
space on which the M(atrix) theory is defined
as a field theory. When the $U$-algebra is
noncommutative, the dual space is a quantum space.
Thus in a sense T duality naturally introduces
the ideas of noncommutative geometry into
M(atrix) theory.

For the compactifications on $\cal{M}/\bG$
with $\cal{M}$ simply connected, we have seen
in the above examples that for a factor of $\Z$
in $\bG$ there is a factor of $S^1$ in the dual space.
(Note that this statement is more general than
the statement that the dual space of a circle is a
circle, because there can be different compactifications
with the same group $\bG$. They lead to different
field theories on the same dual space.) In the
above we also see that for a factor of $\Z_n$ in $\bG$
there is a factor of the dual $\Z_n$ in the dual space.
It would be useful to know more about the correspondence
between the group $\bG$ and the dual space.

\section{Comments and Discussions}\label{Discussion}

Finally we make a few remarks.

To be treated as a fundamental theory by itself,
M(atrix) theory needs to know everything without
consulting string theory or supergravity.
Since the notion of spacetime is from the very beginning
noncommutative in M(atrix) theory,
{\it a priori} one is allowed to consider
compactifications on spaces which are exotic from
a classical point of view. The criterion for
an admissible compactification is only whether
the corresponding generalized gauge theory
on the dual space can make sense.

For compactifications on a classical
$d$-torus, the fundamental group is
commutative and is $d$ dimensional, thus it
results in a $d$-dimensional dual space.
For compactifications on Riemann surfaces of
higher genus, the fundamental group is noncommutative
and therefore the dual space must be a quantum space.

A Riemann surface of genus $g>1$ can be obtained
as a quotient of the Lobachevskian disc which is
simply connected. The quotient conditions are
of the form
\be
U_i^{\dag}Z U_i=\frac{a_i Z+b_i\one}{c_i Z+d_i\one},
\quad i=1,\cdots,2g,
\eq
where $\left(\begin{array}{cc}a_i&b_i\\c_i&d_i\end{array}\right)$
are $SU(1,1)$ matrices and $|Z|<1$.
It is a challenge to find the solution for $Z$.

For two classical compactifications,
it is possible that there is a family of
compactifications on non-classical spaces
with sensible dual theories interpolating them.
Such interpolation may help our understanding
of the various dualities \cite{BD}.

Obviously there are a lot of important issues
we need to clarify before we can proceed further.
If the solution of the quotient conditions gives
us an anomalous gauge theory, what we have
obtained in this paper is only the so-called
untwisted sectors in M(atrix) theory. To view
M(atrix) theory as a fundamental theory, we
also need to learn how to determine the
twisted sectors for anomaly cancellation
without consulting with string theory.
On the other hand, for the consideration
of quantum spaces to be physically relevant,
it is urgent to look for more correspondence
between M(atrix) compactification on quantum
space and the moduli space of M theory
compactification.

\section*{Acknowledgment}

P.M.H. thanks Bruno Zumino for discussion
and Igor Klebanov for hospitality
at Princeton University.
Y.Y.W. thanks Jonathan Bagger for discussion.
This work is supported in part by
U.S. NSF grant PHY-9601277 and PHY-9404057.
Y.S.W. is also supported by a fellowship
from Monnell Foundation.
\appendix

\section{$T^2/\Z_3$} \label{T2Z3}

The quotient conditions for $T^2/\Z_3$ are
\bee
U_1^{\dag}Z U_1&=&Z+\one, \\
U_2^{\dag}Z U_2&=&Z+\tau, \\
U_3^{\dag}Z U_3&=&qZ,
\eqq
where $\tau=q=\exp(2\pi i/3)$ and
$Z=(X_1+iX_2)/R_1$.

The $U$-algebra is given by
\bee
U_1 U_2&=&q_{12}U_2 U_1, \\
U_1 U_3&=&q_{13}U_3 U_1^{\dag}U_2^{\dag}, \\
U_3 U_1&=&q_{31}U_2 U_3, \\
U_3 U_2&=&q_{32}U_1^{\dag}U_2^{\dag}U_3, \\
U_3^3&=&q_3\one, \label{333}
\eqq
where $q_{12}, q_{13}, q_{31}$ are phases and
consistency requires $q_{32}=q_{13}q_{31}^{-1}$.
By rescaling the $U$'s we can set all
the $q$ factors to one except that
$q_{12}$ is still arbitrary.

The Hilbert space $\H$ is
$\{U_1^m U_2^n U_3^s\ra| m,n\in\Z, s=0,1,2\}$.
Define operators $\del_i, \D_s, K$ by
\bee
\del_1 U_1^m U_2^n U_3^s\ra&=&m U_1^m U_2^n U_3^s\ra, \\
\del_2 U_1^m U_2^n U_3^s\ra&=&n U_1^m U_2^n U_3^s\ra, \\
\D_s U_1^m U_2^n U_3^{s'}\ra&=&\d_{ss'}U_1^m U_2^n U_3^{s'}\ra, \\
K U_1^m U_2^n U_3^s\ra&=&U_1^m U_2^n U_3^{s+1}\ra,
\eqq
where $\d_{ss'}=1$ if $s-s'=0$ (mod 3), and vanishes otherwise.

Let $Z\ra=\hA(U)\ra$ and
$\hA(U)=\sum_{mns}\a_{mns}U_1^m U_2^n U_3^s$.
Then
\bee
Z U_1^m U_2^n U_3^s\ra&=&U_1^m U_2^n U_3^s(m+\t n+q^s\hA)\ra \nn\\
&=&(\del_1+\t\del_2+A)U_1^m U_2^n U_3^s\ra,
\eqq
where
\be
A=\sum_{m,n\in\Z; s=0,1,2}\a_{mns}K^s\left(
\sum_{s'=0,1,2}(U_3^{s'}\tU_2 U_3^{-s'})^n(U_3^{s'}\tU_1 U_3^{-s'})^m
q^{s'}\D_{s'}\right),
\eq
where $\tU_1=q_{12}^{-\del_2}U_1$ and $\tU_2=q_{12}^{\del_1}U_2$.
It is not hard to calculate
$U_3\tU_1 U_3^{-1}=\tU_2$,
$U_3^2\tU_1 U_3^{-2}=U_3\tU_2 U_3^{-1}=\tU_1^{-1}\tU_2^{-1}$
and $U_3^2\tU_2 U_3^2=\tU_1$.
The solution of $Z$ is thus
\be \label{ZMNA}
Z=\del_1+\t\del_2+A.
\eq

To put the result in a more amiable form,
let $U_1=\exp(i\s_1)$ and $U_2=\exp(i\s_2)$.
Also let $U_3=P\circ U$, where $U$ is given by (\ref{UV}) for $n=3$
and $P$ is an algebraic operation defined by
\bee
&P\s_1 P^{-1}=\s_2, & P^2\s_1 P^{-2}=-\s_1-\s_2, \\
&P\s_2 P^{-1}=-\s_1-\s_2, & P^2\s_2 P^{-2}=\s_1.
\eqq
Then it is easy to see that (\ref{ZMNA}) can be rewritten as
\be
Z=(-i\frac{\del}{\del\s_1}-i\t\frac{\del}{\del\s_2})\one+A(\s_1,\s_2),
\eq
where $\one$ is the $3\times 3$ unit matrix and
$A$ is a $3\times 3$ matrix of functions of $(\s_1,\s_2)$ satisfying
\be
A_{i-1,j-1}(\s_1,\s_2)=qA_{ij}(\s_2,-\s_1-\s_2+\pi/3),
\eq
where the indices are defined modulo $3$.
The dual space is again $T^2/\Z_3$.


The rest of the quotient conditions are fixed by
the Lagrangian (\ref{Lag}) to be
\bee
U^{\dag}X_{\mu}U&=&X_{\mu}, \quad \mu=0,3,\cdots,9, \\
U^{\dag}\Psi U&=&\Lam_3\Psi,
\eqq
where $\Lam_3=\exp(-\pi\G^{12}/3)$.
Because $\Lam_3^3=-\one$,
strictly speaking eq.(\ref{333}) should be
replaced by $U_3^3=(-1)^F$,
where $F$ is the fermion number operator.
All the SUSY is broken in this case.

\section{Finite Cylinder} \label{cylinder}

Matrix compactification on the orientifold $S^1\times S^1/\Z_2$
is related to the heterotic string theory \cite{Motl,het}.
The quotient conditions are \cite{Motl,het}
\footnote{In general there can be an additional term of
$2k\pi R_1$ for any integer $k$ in (\ref{X1}),
but it can be absorbed in a shift of $X_1$ by
$X_1\rightarrow X_1+k\pi R_1$.}
\bee
U_i^{\dag}X_j U_i&=&X_j+2\pi\d_{ij}R_j, \quad i,j=1,2, \label{T2'}\\
U_3^{\dag}X_1 U_3&=&-X_1^*, \label{X1'}\\
U_3^{\dag}X_2 U_3&=&X_2^*. \label{X2'}
\eqq

The $U$-algebra is
\bee
U_1 U_2&=&q_{12}U_2 U_1, \label{U12'}\\
U_1 U_3&=&q_{13}U_3 U_1^T, \label{U13'}\\
U_2 U_3&=&q_{23}U_3 U_2^*, \label{U23'}\\
U_3 U_3^*&=&q_3\one. \label{U33'}
\eqq
Consistency of the $U$-algebra imposes constraints
on the parameters $q_{ij}$'s.
Taking the complex conjugation of (\ref{U33'}),
we find $q_3=\pm 1$.
Eq. (\ref{U33'}) and the transpose of (\ref{U13'})
imply that $q_{13}=\pm 1$.
Rescaling $U_2$ can give $q_{23}=1$.
The $U$-algebra is therefore parametrized by a phase $q=q_{12}$,
$q_{13}=\pm 1$ and $q_3=\pm 1$.
For $q=q_{13}=q_3=1$ we get the same algebra as in \cite{Motl,het}.

The Hilbert space is
$\H=\{U_1^m U_2^n(U_3 C)^s\ra| m,n\in\Z, s=0,1\}$.
Define $\del_i, K$ and $\eps$ by
\bee
\del_i U_1^{m_1}U_2^{m_2}(U_3 C)^s\ra&=&
m_i U_1^{m_1}U_2^{m_2}(U_3 C)^s\ra, \\
K U_1^m U_2^n(U_3 C)^s\ra&=&U_1^m U_2^n(U_3 C)^{s+1}\ra, \\
\eps U_1^m U_2^n(U_3 C)^s\ra&=&(-1)^s U_1^m U_2^n(U_3 C)^s\ra.
\eqq

To follow Zumino's prescription, we consider
\bee
X_i U_1^{m_1} U_2^{m_2}\ra&=&
U_1^{m_1} U_2^{m_2}(2\pi m_i R_i+\hA_i(U_1,U_2,U_3)\ra \nn\\
&=&(2\pi R_i\del_i+A_i(\tU_1,\tU_2,K))U_1^{m_1} U_2^{m_2}\ra.
\eqq
If $\hA_i=\sum_{mns}\a^{i}_{mns}U_1^m U_2^n(U_3 C)^s$
then $A_i=\sum_{mns}\a^{i}_{mns}\tU_2^n \tU_1^m K^s$,
where $\tU_1=q^{-\del_2}U_1$, $\tU_2=q^{\del_1} U_2$.
Similarly,
\bee
X_i U_1^{m_1} U_2^{m_2} U_3 C\ra&=&
U_1^{m_1} U_2^{m_2} U_3 C(2\pi m_i R_i
+(-1)^i\hA_i(U_1,U_2,U_3)\ra \nn\\
&=&(2\pi R_i\del_i+(-1)^i A^*_i(q_{13}\tU_1^{-1},\tU_2,K))
U_1^{m_1} U_2^{m_2} U_3\ra.
\eqq
Therefore we get
\be \label{cylinder-X}
X_i=2\pi R_i\del_i\,+\,\frac{1}{2}A_i(\tU_1,\tU_2,K)(1+\eps)
\,+\,(-1)^{i}\frac{1}{2}B_i(\tU_1,\tU_2,K)(1-\eps),
\eq
where $B_i(\s_1,\s_2,K)=A^*_i(\s_1-h_{13},-\s_2,K)$
with $\tU_1=e^{i\s_1}$, $\tU_2=e^{i\s_2}$
and $q_{13}=e^{ih_{13}}$ ($h_{13}=0, \pi$).
The fundamental region on which the gauge field
can be freely assigned is a dual cylinder:
$\s_1\in[0,2\pi), \s_2\in[0,\pi]$ for $q_{13}=1$.
For $q_{13}=-1$ it is a dual Klein bottle.

Let $A_i=A_{i0}(\tU_1,\tU_2)+A_{i1}(\tU_1,\tU_2)K$
and similarly for $B_i$.
The Hermiticity of $A_i$ implies that
\be
A_{i0}^{\dag}=A_{i0}, \quad B_{i0}^{\dag}=B_{i0},
\quad A_{i1}^{\dag}=(-1)^i q_3 B_{i1}.
\eq

Clearly, $\del_1, \del_2$ are derivatives on the dual space.
In fact $K$ can also be viewed as a function on $\Z_2$
and $\eps$ as the derivative on $\Z_2$
in the sense of noncommutative geometry \cite{Con}.
Hence the dual quantum space is the product of
the dual cylinder with $\Z_2$.
Furthermore, the form of $X$ resembles the covariant derivative
on the dual quantum space as defined in \cite{Con,CL}.
A similar construction was used for rewriting the standard model
as a gauge theory on a noncommutative space \cite{CL}.


The algebra on the $\Z_2$ factor of dual space
can be represented by Pauli matrices.
For instance, $K=\t_1$ and $\eps=\t_3$ for $q_3=1$.
{}From (\ref{cylinder-X}),
$X_i=-i 2\pi R_i\frac{\del}{\del\s_i}+{\cal A}_i(\s_1,\s_2)$,
where
\be
{\cal A}_i=\left(\begin{array}{cc}
A_{i0} & (-1)^i B_{i1} \\ A_{i1} & (-1)^i B_{i0} \end{array}\right)
\eq
is a Hermitian matrix.
Each entry of the $2\times 2$ matrices
is an $N\times N$ matrix.

The quotient conditions for other coordinates
for the compactification on a cylinder are \cite{Motl,het}
\bee
U_i^{\dag} A_0 U_i=A_0, & U_3^{\dag} A_0 U_3=-A_0^*, \\
U_i^{\dag} X_a U_i=X_a, & U_3^{\dag} X_a U_3=X_a^*,  \\
U_i^{\dag}\Psi U_i=\Psi, & U_3^{\dag}\Psi U_3=\G_{01}\Psi^*,
\eqq
where $i=1,2$, and $a=3,\cdots,9$.
The M(atrix) theory on a cylinder is related to
the heterotic string theory \cite{Motl,het}.
It is a gauge theory with the gauge group $U(2N)$
in the bulk of the dual cylinder but with
the gauge group $O(2N)$ ($q_{13}=1$) or $USp(2N)$ ($q_{13}=-1$)
on the boundary \cite{Motl}.
One half of the dynamical SUSY is preserved.

\section{M\"{o}bius Strip} \label{Mobius}

The quotient conditions for a M\"{o}bius strip \cite{Zwa,KR2}
are (\ref{T2'}) and
\bee
U_3^{\dag}X_1 U_3&=&X_2^*, \label{X12}\\
U_3^{\dag}X_2 U_3&=&X_1^*. \label{X21}
\eqq

The $U$-algebra is
\bee \label{Mobius-U}
U_1 U_2&=&q_{12}U_2 U_1, \\
U_1 U_3&=&q_{13}U_3 U_2^*, \label{U1232}\\
U_2 U_3&=&q_{23}U_3 U_1^*, \label{U21}\\
U_3 U_3^*&=&q_3\one.
\eqq
Considerations similar to those in the previous sections
lead to $q_3=\pm 1$ and $q_{13}=q_{23}=1$.
The phase $q_{12}=q$ and $q_3=\pm 1$ label two
one-parameter families of compactifications.

The Hilbert space and the operators $\del_i, K, \eps$
can be defined similarly as in the previous section.
We get the solution for $X_1, X_2$ as
\be
X_i=2\pi R_i\del_i + \frac{1}{2}A_i(\tU_1,\tU_2,K)(1+\eps)
+ \frac{1}{2}B_i(\tU_1,\tU_2,K)(1-\eps),
\eq
where the $A$'s and $B$'s are functions of
$(\tU_1,\tU_2)=(q^{-\del_2}U_1,q^{\del_1}U_2)=(e^{i\s_1},e^{i\s_2})$.
It is
\be
A_i(-\s_2,-\s_1)=B_j^*(\s_1,\s_2),
\eq
where $(i,j)=(1,2)$ or $(2,1)$.
{}From (\ref{X12}), (\ref{X21}), (\ref{U1232}) and (\ref{U21}),
the fundamental region is the dual M\"{o}bius strip
and the compactified M(atrix) theory is a field theory
on the dual M\"{o}bius strip.

The quotient conditions for $A_0$ and $X_a (a=3,\cdots,9$) are
the same as those for a cylinder.
Those for $\Psi$ can also be obtained:
\bee
U_i^{\dag}\Psi U_i&=&\Psi, \\
U_3^{\dag}\Psi U_3&=&\G_{\perp}\Psi^*,
\eqq
where $\G_{\perp}=\frac{1}{\sqrt{2}}\G_0(\G_1-\G_2)$.
One half of the dynamical SUSY is preserved.

\end{document}